\documentclass{ws-rv961x669}
\pdfoutput=1
\usepackage{ws-rv-van}     
\usepackage{ws-rv-thm}     
\usepackage{subfigure}     
\usepackage{slashed}
\makeindex

\begin{document}

\chapter[A de-gauging approach to physics beyond the Standard Model ]{A de-gauging approach to  physics beyond the Standard Model }\label{ra_ch1}

\author[CHI XIONG]{CHI XIONG \footnote{E-mail: xiongchi@ntu.edu.sg}}

\address{Institute of Advanced Studies, Nanyang Technological University,\\
60 Nanyang View, Singapore 639673, Singapore.}

\begin{abstract}
By studying the t-J model for superconductivity, the Pati-Salam model and the Haplon model for particle unifications, we extract their common feature which is the spin-charge separation of fermions. This becomes a de-gauging process for charged fermions by considering them as bound states of a neutral fermion and charged or neutral bosons.  We present a few examples including the weak-charge-spin separation for the leptons in the Standard Model. Some fundamental fermions can be obtained by continuing this de-gauging process for different kinds of charges. Finally the binding forces of the bound states might be provided by interactions related to spacetime symmetries such as supersymmetry.       
\end{abstract}



\body


\section{Introduction}
~~~~~The new physics beyond the Standard Model, albeit currently unknown in experiments, should be able to address the problem of neutrino mass and provide interpretations for other problems such as the naturalness problem, the mass hierarchy problem and the repetitive family structure. More ambitious goals, like what grand unification theories (GUTs) are aimed to achieve, are usually set in the framework of gauge theories with spontaneously broken symmetry. Fundamental or composite Higgs bosons are needed in describing the physics vacua and the symmetry-breaking patterns. It is generally speculated that gauge-theory side of leptons and quarks in the Standard Model is ``fixed" to some extent, while the Higgs sector still has some room to be improved. Given that in the study of superconductivity the Bardeen-Cooper-Schrieffer (BCS) theory surpasses the phenomenological Ginzburg-Landau theory,  we agree with the second part of this speculation --- the Higgs boson could be composite; there might be more Higgs fields and etc --- but not the first part: the leptons, and probably quarks, might be composite as well. This is not new from the ``preon model" point of view \cite{Collins}. What is new is that we propose a ``{\it de-gauging}" procedure, in the spirit of the {\it spin-charge separation} from the study of high-temperature superconductivity in condensed matter physics. The idea is that we decompose the leptons into Higgs fields and some neutral fermions, called ``{\it dark fermions}" in Ref. \cite{XC-16}, and leptons can be considered as bound states of these component fields. Therefore the leptons masses and the Yukawa couplings, as well as the family structure, might come from the dynamics of the bound states. As this de-gauging procedure goes on, the fermionic components are becoming ``darker" in the sense that they are carrying less and less charges. Finally only interactions with respect to the spacetime symmetry, like gravity and supersymmetric interactions, are left for providing binding forces of the bound states. Additionally the dark fermions we obtained during the de-gauging course, are proper candidates for dark matter. 

This article is organized as follows: In Sect. 2 we will give a brief introduction to the spin-charge separation in condensed matter physics; We then show in Sect. 3 that in particle physics, some GUTs (Pati-Salam) and preon models do have a spin-charge separation pattern; As examples, we will study the de-gauging of the sigma model and the vectorial gauge theories in Sect. 4; The spin-charge separation of the Standard Model leptons is done in Sect. 5 and discussions and conclusions are given Sect. 6.

\section{Spin-charge separation from condensed matter physics}

~~~~~Compared to the BCS theory for superconductivity, spin-charge separation might not be a familiar topic for particle physicists. In condensed matter physics, spin-charge separation is associated with the resonating valence bond (RVB) theory \cite{Anderson} and describes electrons in some materials as ``bound states" of spinon and chargon (or holons), which carry the spin and charge of electrons respectively. Under certain conditions, e.g. in the high-temperature cuprate superconductors, the ``composite" electrons can have a deconfinement phase and the spinon and chargon becomes independent excitations or ``particles". Many elaborations of this idea followed in the studies of high $T_c$ superconductors (see Ref. \cite{Wen:2006} for a review).  Figure 1 and Figure 2 illustrate the BCS theory and the spin-charge separation mechanism in two cartoons, respectively. 

\begin{figure}[!t]
\begin{center}
\includegraphics[width=3.2in]{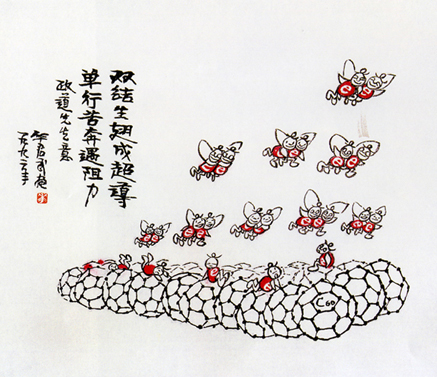}
\caption{A cartoon for the BCS mechanism for superconductivity, by Junwu Hua and T. D. Lee for the Conference on "High-temperature conductors
and $C_{60}$ family". It illustrates that the interaction between the electrons and the vibrating crystal lattice brings an attractive force between electrons, hence making it possible for electrons to become correlated and form pairs (Cooper pairs), ``flying" though the material without impedance.}
\end{center}
\end{figure}

\begin{figure}[!t]
\begin{center}
\includegraphics[width=3.2in]{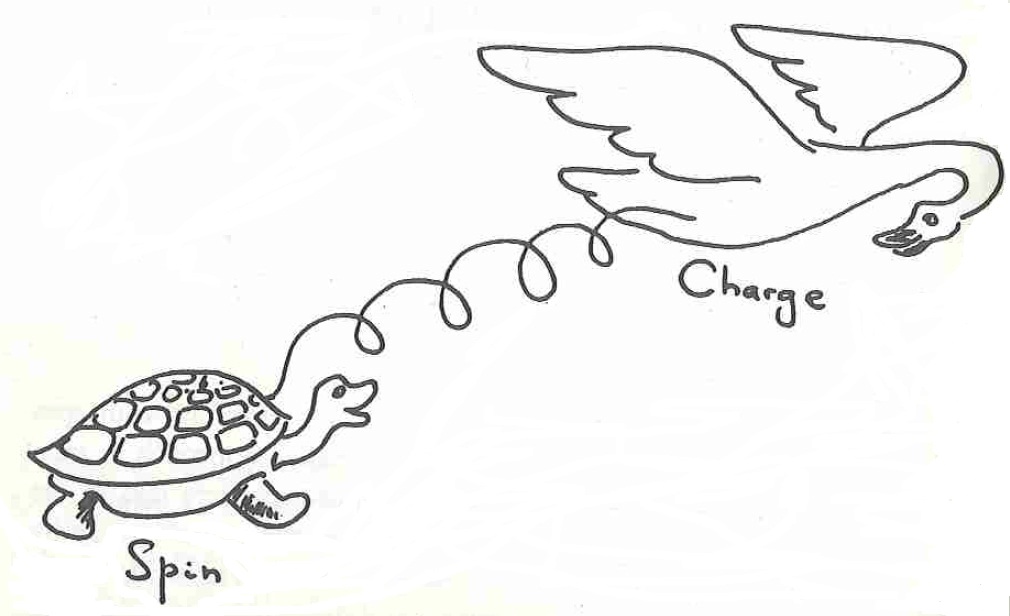}
\caption{A cartoon illustrating spin-charge separation (from Ref. \cite{Tsvelik}, courtesy of A. Tsvelik). For one-dimensional interacting systems of fermions with spin, collective excitations, instead of renormalized electrons, are the coherent excitations in the bosonization approach and there are two branches of collective excitations, the charge density waves and the spin density waves, which carry charge $\pm e$, spin 0 and charge 0, spin $1/2$, respectively. These two branches have different symmetries and hence different spectra. To some extremes one branch has a gap and the other does not, so the electrons carrying both spin and charge quantum numbers cannot propagate coherently. The two branches in electrons under such circumstances tend to separate from each other (see Ref.\cite{Tsvelik} for details).}
\end{center}
\end{figure}

There are experimental observations and computer simulations supporting the idea of spin-charge separation --- The first direct observations of spinons and holons was reported in Ref. \cite{Shen:2006}; Simulations on spin-charge separation with quantum computing methods were performed in Ref. \cite{Kwek:2011}. It is more convenient for us to demonstrate this concept by taking the slave-boson formalism in the t-J model as an example \cite{Barnes:1976}.  It is well-known that the low-energy physics of the high-temperature cuprates can be described by the t-J model
\begin{equation}
H = \sum_{ij} J ( {\bf S}_i \cdot {\bf S}_j - \frac{1}{4} n_i n_j ) - \sum_{ij, \sigma} t_{ij} ( c^{\dagger}_{i \sigma} c_{j \sigma} + h.c.),
\end{equation}
where $c^{\dagger}_{i \sigma}, c_{i \sigma}$ are the projected electron operators with constraint $\sum_\sigma c^{\dagger}_{i \sigma} c_{i \sigma} \leqslant 1$, and $t_{ij} = t, t', t''$ label the nearest-, second nearest- and their-nearest-neighbor pairs, respectively.  With double occupancy excluded one can write
\begin{equation} \label{cfb}
 c^{\dagger}_{i \sigma} = f^{\dagger}_{i \sigma} b_i 
\end{equation}
where the operators $f^{\dagger}_{i \sigma}$ creates a chargeless spin-1/2 fermion state -- spinon, and $b_i $ creates a charged spin-0 boson state -- holon, respectively. Note that for the (anti)commutator relations to work out correctly a constraint:
\begin{equation} \label{constraint}
f^{\dagger}_{i \uparrow} f_{i \uparrow} +  f^{\dagger}_{i \downarrow} f_{i \downarrow} + b^{\dagger}_i b_i =1
\end{equation}
must be satisfied. This reflects the material or environmental dependence of the decomposition (\ref{cfb}). What's more, a U(1) gauge symmetry 
\begin{equation}
b_i \rightarrow e^{i \theta} b_i, ~~ f_{i \sigma} \rightarrow e^{i \theta} f_{i \sigma}
\end{equation}
emerges. The d-wave high-$T_c$ superconducting phase appears when holons condense $\left\langle b_i^{\dagger} b_i \right\rangle \neq 0.$  It should be emphasized that, as Eq. (\ref{constraint}) suggests, to apply spin-charge separation in condensed matter systems we need a finite-density environment. An isolated electron cannot split into its spin and charge components. Besides the high-$T_c$ cuprate superconductors in the condensed matter physics, the early universe and the inner cores of the compact stars may provided such a finite-density environment in cosmology and astrophysics.

In particle physics, however, there exists a different interpretation to the decomposition in (\ref{cfb}) --- it may indicate that the particle state $c$ is a bound state of component fields $f$ and $b$. Composite models are quite common in physics and one may ask what is special about the decomposition in  (\ref{cfb}). This will be answered in the next two sections.

\section{Spin-charge separation hidden in the Pati-Salam model and the haplon model}
~~~~~It is interesting that the decomposition pattern (\ref{cfb}) has already been put forward in some unification models in particle physics. To our knowledge the first one seems to be the Pati-Salam model \cite{Pati-Salam-74}, in which the lepton number can be considered as the fourth ``color" \cite{Pati-Salam-74}, based on the gauge group $SU(2)_L \times SU(2)_R \times SU(4)_c$. (Note that this group can be embedded into a larger gauge group $SO(10)$ \cite{Fritzsch:1974} and the observations and arguments below should work as well in the $SO(10)$ or even larger group cases. Right-handed neutrinos are introduced in these models \cite{Pati-Salam-74, Fritzsch:1974} and the usual color group $SU(3)_c$ is extended to the group $SU(4)_c$. With modern notations
\begin{equation}  \label{FB}
\Psi_{L, R} = \left( \begin{array}{cccc}
u_1 & u_2 & u_3 & u_4 = \nu_e \\ 
d_1 & d_2 & d_3 & d_4 = e^{-} \\ 
c_1 & c_2 & c_3 & c_4 = \nu_{\mu} \\ 
s_1 & s_2 & s_3 & s_4 =  \mu^{-} 
\end{array} \right)_{L, R} = \left( \begin{array}{c}
\mathcal{F}_1 \\ 
\mathcal{F}_2 \\ 
\mathcal{F}_3 \\ 
\mathcal{F}_4
\end{array}  \right)_{L, R} \otimes (\mathcal{B}_1, \mathcal{B}_2,  \mathcal{B}_3,  \mathcal{B}_4)
\end{equation}
where $\mathcal{F}$ are spinors carrying the spin of $\Psi$ while $\mathcal{B}$ are scalars carrying the {\it color} charge of $\Psi$. As explicitly pointed out in Ref. \cite{Pati-Salam-74}, it is attractive to consider the components $(\mathcal{F}, \mathcal{B})$ in the decomposition (\ref{FB}) as  {\it fundamental fields} and $\Psi$ as {\it composite} ones. Note that this actually suggests a composite model for quarks and leptons, which can be considered as one of those ``preon" models (preons are hypothetical constituents of the elementary particles in the Standard Model) \cite{Collins} .  

Similar pattern can be found in another preon model called the ``Haplon" model \cite{Fritzsch:1981}, which consider quarks and leptons as bound states of some preons called Haplons. In the Haplon model the preons are in the fundamental representation $N$ or $\bar{N}$ of an $SU(N)$ hypercolor gauge group. They are a weak SU(2) doublet of colorless fermions $(\alpha, \beta)$, and a quartet of scalars $(x^i, y), i = 1, 2, 3$ and $y$ carries the fourth color, i.e. lepton number, thus leading to an $SU(4)$ symmetry. The quantum number assignment are given in Table \ref{Table1} \cite{Collins}. Notice that there are two different ways of assigning the color charge to the Haplons, as shown in the last two columns of Table \ref{Table1}. The original scheme in Ref.  \cite{Fritzsch:1981} takes the alternative one (the last column), which does not have a spin-charge separation while the other scheme (the second last column) does. Therefore we choose the color assignment in the second last column for spin-charge separation purpose. The first generation of fermions 
\begin{equation} \label{haplon}
\nu = (\alpha y), ~e^{-} = (\beta y), ~ u = (\alpha x), ~d = (\beta x). 
\end{equation}
which has the same spin-charge separation pattern $\Psi = \mathcal{F} \mathcal{B}$  as in Eqns. (\ref{cfb}) and (\ref{FB}). The weak vector bosons (spin 1 state $\uparrow \uparrow$) and Higgs scalars (spin 0 state $\uparrow \downarrow)$ can also be composite ones \cite{Fritzsch:1981}
\begin{equation}
W^{+} = (\alpha\bar{\beta}), ~W^{-} = (\bar{\alpha} \beta), ~ W^3 = \frac{1}{\sqrt{2}} (\alpha\bar{\alpha} + \beta\bar{\beta}), ~Y^0  =  \frac{1}{\sqrt{2}} (\alpha\bar{\alpha} - \beta\bar{\beta})
\end{equation}
However,  we will not address here the decomposition of gauge field. The spin-charge separation of non-abelian gauge fields is a more complicated issue (see for example Refs. \cite{Niemi:2005, Faddeev:2006}).

\begin{table}[h]
\tbl{The Haplon model.}
{\begin{tabular}{c c c c c c}
\hline 
Halplon & Spin ($\hbar $) & Charge ($e$)  & Hypercolor & Color & Color (alternative)$^{\text a}$ \\ 
\hline 
$\alpha$ & $\frac{1}{2}$ & $\frac{1}{2}$ & $N$ & 1 & $\bar{3}$ \\ 
$\beta$ & $\frac{1}{2}$ & $-\frac{1}{2}$ & $N$ & 1 & $\bar{3}$ \\ 
$x$ & 0 & $\frac{1}{6}$ & $\bar{N}$ & $3$ & $\bar{3}$ \\ 
$y$ & 0 & $-\frac{1}{2}$ & $\bar{N}$ & $1$ & $3$ \\ 
\hline 
\end{tabular}}
\begin{tabnote}
$^{\text a}$ This is the original choice in Ref.\cite{Fritzsch:1981}. For spin-charge separation purpose we choose the color assignment in the second last column. 
\end{tabnote}\label{Table1}
\end{table} 

\section{Spinon and chargon: some examples}

~~~~~From Eqns (\ref{cfb}, \ref{FB}, \ref{haplon}), the $t-J$ model, the Pati-Salam model and the Haplon model  tell us that a general spin-charge separation can be written as 
\begin{equation} \label{gFB}
 \Psi = \mathcal{F} \mathcal{B}
\end{equation}
Extra constraint(s) like Eq. (\ref{constraint}) might be needed to form the correct (anti) commutation relations for the operators, and to match the degrees of freedom before and after the spin-charge separation. How to do a spin-charge separation in field theories?  First we consider the sigma model as a simple example. The Lagrangian of the (linear) sigma model with $SU(2)_L \times SU(2)_R$ symmetry reads
\begin{equation}
\mathcal{L}_{\sigma} = \bar{\psi}_L i \slashed{\partial} \psi_L + \bar{\psi}_R i \slashed{\partial} \psi_R + \frac{1}{4} \textrm{Tr} \,(\partial \Sigma \cdot \partial \Sigma^{\dagger}) - V(\Sigma, \Sigma^{\dagger}) - y ( \bar{\psi}_L \Sigma \psi_R +  \bar{\psi}_R \Sigma^{\dagger} \psi_L ),
\end{equation}
where $V = \mu^2/4 ~ \textrm{Tr} (\Sigma^{\dagger} \Sigma) - \lambda/16 ~ [\textrm{Tr} (\Sigma^{\dagger} \Sigma)]^2 $. The $SU(2)_L \times SU(2)_R$ transformations of the fields $\psi_{L,R}$ and $\Sigma$ are
\begin{equation}
\psi_{L,R} \rightarrow \mathcal{A}_{L, R} \, \psi_{L,R}, ~\Sigma \rightarrow \mathcal{A}_L \Sigma \mathcal{A}_R^{\dagger}, ~~~ \mathcal{A}_{L, R} \in SU(2)_{L,R}
\end{equation}
The matrix field $\Sigma$ can be parametrized by either the fields  $\sigma$ and $\vec{\pi}$ or the polar variables $\eta$ and $\vec{\zeta}$ 
\begin{equation} \label{scs_Sigma}
\Sigma = \sigma (x) + i \vec{\tau} \cdot \vec{\pi} (x) =  [v + \eta(x)] U(x), ~~~U= e^{i \vec{\tau} \cdot \vec{\zeta}(x)/v },
\end{equation}
where $v = \sqrt{\mu^2/\lambda}$ is the vacuum expectation value of $\sigma$. In terms of $\eta$ and $U$, the $SU(2)_L \times SU(2)_R$ transformation laws become 
\begin{equation}
\eta \rightarrow \eta, ~~ U \rightarrow \mathcal{A}_L U \mathcal{A}_R^{\dagger}. 
\end{equation} 
Therefore the $U$ field inherits the transformation property of $\Sigma$ while the $\eta$ field is invariant.  From spin-charge separation point of view,  we consider the $\eta$ field as a ``spinon" $\mathcal{F}$, albeit a real scalar with $0$-spin in this case, and the $U$ field as the chargon $\mathcal{B}$, respectively. Also the second parametrization in (\ref{scs_Sigma}) is the spin-charge separation for the $\Sigma$ field.   

If a free, massless Dirac fermion $\Psi_D$ is a bound state of a spinon and a chargon $(\mathcal{F} \mathcal{B})$, then from the wave function point of view the component fields $\mathcal{F}$ and $\mathcal{B}$ would be coupled to each other as can be seen from plugging $\Psi_D =\mathcal{F} \mathcal{B} $ into its Lagrangian, in this case just the kinetic term  $ i \bar{\Psi}_D \gamma^\mu \overleftrightarrow{\partial}_{\mu} \Psi_D $, 
\begin{equation} \label{JJ}
~J_{\tiny{\textrm{spinon}}} \cdot J_{\tiny{\textrm{chargon}}} \sim  \bar{\mathcal{F}} \gamma^\mu \mathcal{F} \left(  \mathcal{B}^{\ast}\partial_{\mu}\mathcal{B} -\mathcal{B}\partial_{\mu}\mathcal{B}^{\ast}\right) \sim J^\mu_{\tiny{\textrm{spinon}}} \partial_\mu \xi
\end{equation}
where the spinon current and the chargon current are defined as 
\begin{equation}
J^\mu_{\tiny{\textrm{spinon}}} \equiv  \bar{\mathcal{F}} \gamma^\mu \mathcal{F}, ~~J^{\mu}_{\tiny{\textrm{chargon}}} \equiv 1/(2i) \left(  \mathcal{B}^{\ast}\partial^{\mu}\mathcal{B} -\mathcal{B}\partial^{\mu}\mathcal{B}^{\ast}\right) \sim \partial^\mu \xi ,
\end{equation}
respectively and the field $\xi$ is the phase of the chargon field  $\mathcal{B}$. The coupling (\ref{JJ}) has the usual form of derivative coupling of Nambu-Goldstone boson to some external current.

We then consider vector-like theories which the spinons $\mathcal{F}$ will be Majorana type. Starting with an Abelian gauge theory for a Dirac fermion $\Psi_D$ and a gauge field $A_\mu$
\begin{equation}
\mathcal{L}_{V} = \bar{\Psi}_D ( i \slashed{\partial} + e \slashed{A} ) \Psi_D - \frac{1}{4} ~F^2. 
\end{equation}
Like the complex scalar field $\Sigma$, the Dirac fermions can be considered as a combination of two Majorana fermions $\Psi_D  = \Psi_M^1 + i \Psi_M^2$, and one may wonder what its polar form or exponential parametrization would be. The spin-charge separation suggests $\Psi = \mathcal{F} \mathcal{B} = \Psi_M \mathcal{B} $, where the ``spinon" $\mathcal{F}=\Psi_M$ is a Majorana spinor. The coupling $A_\mu J^\mu$ then vanishes as $ J^\mu \propto \overline{\Psi}_M \gamma^\mu \Psi_M= 0$,  hence the Majorana spinon $ \Psi_M$ decouples from the gauge field $A_\mu$. The interaction of $\Psi_M$ to the chargon $\mathcal{B}$ vanishes as well since it is proportional to $  ~\partial_\mu \xi ~ \overline{\Psi}_M \gamma^\mu \Psi_M$ from Eqn. (\ref{JJ}). Thus we obtain a complete spin-charge separation (or de-gauging) of a Dirac fermion for this vectorial theory, if the spinon is taken to be the Majorana-type.

\section{De-gauging leptons in the Standard Model}

~~~~~In Ref. \cite{XC-15} we considered whether neutrinos be considered as spinons with respect to the electric-charge-spin separation. Since chirality is involved, and in the (3+1)-dimensional Minkowski spacetime spinors cannot satisfy both Weyl and Majorana conditions, the spin-charge separation or the de-gauging process will become more complicated for chiral theories, like the electroweak theory in the Standard Model. If spin-charge separation can be realized in both electric and color charge cases, it should be realized for the weak-charge case as well and we will show how this could be proceeded. Let us push the idea to extremes --- what about if we separate all the charges of quarks and leptons from their spins? We will obtain some fundamental fermions interacting with each other via forces which are related to the spacetime symmetries, such as gravity and supersymmetric interactions.  Continuing this spin-charge separation might happen step by step (the de-gauging process), we obtain dark fermions \cite{XC-16} which might be the proper candidates for dark matter.

It turns out that the spin-charge separation has to be performed in an asymmetric way (with respect to the left-right symmetry) \cite{XC-16}.  As the neutrinos and electrons form an $SU(2)_L$ doublet in the Standard Model, we consider their weak-charge-spin separation together. Let us start with the first generation. (The family-structure problem will be discussed in Sect. 6).  Noticing that left-handed lepton doublet $l_L \equiv ( \nu, e^-)^T_L$ and the tilde Higgs doublet $\tilde{\Phi}$ both transform as $(1, 2, -1/2)$ under the symmetry group $SU(3)_C \times SU(2)_L \times U(1)_Y$, a wild guess would be
\begin{equation} \label{guess}
 \left( \begin{array}{c}
\nu \\ 
e^-
\end{array} \right)_L \sim  \tilde{\Phi} \otimes \mathcal{F}_L = \left( \begin{array}{c}
~\phi^{0*} \mathcal{F}_L \\ 
- \phi^- \mathcal{F}_L 
\end{array}  \right)
\end{equation}
where the neutral spinor $\mathcal{F}_L$ plays the role of spinon, and the Higgs fields are the chargons.  For the right-handed lepton singlets of $SU(2)_L$, we need extra scalar fields $\chi^-$  and $\chi^0$ for their chargon parts
\begin{equation}
e^{-}_R \sim \chi^- \mathcal{F}_R, ~~~\nu_R  \sim \chi^0 \mathcal{F}_R. 
\end{equation}
where the spinor $\mathcal{F}_R$ is another spinon. Note that even from the composite model point of view, this is different from the Haplon model in comparison with (\ref{haplon}). Now we give a derivation to find the precise form of (\ref{guess}).  For compact gauge groups it is always possible to choose the unitary gauge and write the Higgs doublet in a polar form similar to (\ref{scs_Sigma})
\begin{equation} \label{Ugauge}
\Phi (x)  = U^{-1} (\xi) \left( \begin{array}{c}
0 \\ 
\frac{h(x)}{\sqrt{2}}
\end{array}  \right), ~~~~~~U(\xi) = e^{-i \vec{\tau} \cdot \vec{\xi}/v}.
\end{equation} 
As mentioned in the sigma-model example, this is the spin-charge separation form for the Higgs doublet and the field $h(x)$ plays the role of spin-0 spinon.  With $U(\xi)$ a field redefinition for the lepton doublet can be made,
\begin{equation} \label{redef}
l_L = \left( \begin{array}{c}
\nu \\ 
e^{-}
\end{array}  \right)_L = U^{-1} \left( \begin{array}{c}
\nu^{'} \\ 
e^{'-}
\end{array}  \right)_L, ~~~~~~e^{-}_R = e^{' - }_R, ~~\nu_{R} = \nu^{'}_{R}. 
\end{equation}
By a straightforward calculation we obtain  
\begin{equation}
\left( \begin{array}{c}
\nu \\ 
e^{-}
\end{array}  \right)_L = \frac{1}{h} \left( \begin{array}{c}
\phi^{0*}\nu^{'} + \phi^{+} e^{'-} \\ 
- \phi^{-} \nu^{'} + \phi^{0} e^{'-}
\end{array}  \right)_L  
\end{equation}
which can be rewritten in a more compact form
\begin{equation}
l_L = \frac{1}{h} \, (\tilde{\Phi} \otimes  \nu^{'}_L + \Phi \otimes e^{'-}_L)
\end{equation}
Noticing that partially de-gauged fields $ \nu^{'}_L$ and $e^{'-}_L$ now has the same quantum number as the $ \nu^{'}_R$ and $e^{'-}_R$, respectively, one may realize that the separation of the weak charge makes the right-handed and left-handed fermions symmetric, as it should be. Therefore let us de-gauge the electron fields  $e^{'-}_L$  and  $e^{'-}_R$ in a left-right symmetric way
\begin{equation} \label{scsEM}
e^{'-}_L = \chi^{-}  \,\mathcal{F}_L~, ~~~~~~~ e^{'-}_R =  \chi^{-} \, \mathcal{F}_R~,
\end{equation}
where the fermion fields $\mathcal{F}_L, \mathcal{F}_R$ are the dark fermions we expect, and the extra scalar $\chi^{-}$ has been made dimensionless by rescaling. For the neutrino fields $ \nu^{'}_L$ and $ \nu^{'}_R$, we assume that they have the {\it same} spinon part as the electron fields, (this is, of course, an assumption which leads to minimum number of dark fermions) 
\begin{equation}
\nu^{'}_L =  \chi^{0} \,\mathcal{F}_L~, ~~~~~~~ \nu^{'}_R =  \chi^{0} \, \mathcal{F}_R~,
\end{equation}
therefore the spin-charge separation for the original fields of the electron and electron neutrino is
\begin{eqnarray}
\nu_L &=& \frac{1}{h} \left( \phi^{0*} + \phi^{+} \chi^{-}  \right) \, \mathcal{F}_L~, ~~~~~~\nu_R =  \chi^{0} \, \mathcal{F}_R \cr
e^{-}_L &=& \frac{1}{h} \left( - \phi^{-} + \phi^{0} \chi^{-}  \right) \, \mathcal{F}_L~, ~~~~~ e^{-}_R = \chi^{-} \, \mathcal{F}_R
\end{eqnarray} 
This is the main result obtained in Ref. \cite{XC-16}. 
Note that we have introduced extra scalars $\chi^-, \chi^0$ for the electric-charge-spin separation. These scalars are important in distinguishing electrons and neutrinos since we have assumed that the spinon part of electrons and neutrinos should be the same. However, it depends on a detailed study of the binding forces and dynamics of the bound states and we leave it to future investigations.

\section{Discussions and conclusions}

Our study suggests that both the (weak) charges of the leptons, just like their masses,  are provided by Higgs fields. \footnote{Note that to acquire electric charge some new Higgs fields are needed.}  We have shown the de-gauging of the first generation leptons. How about the other generations? A naive generalization would be that we introduce spinons or dark fermions for the second and third generation, respectively. But this is again our requirement that the number of spinons should be kept minimum.  Besides, this does not fully take the advantage of the bound states. For example, the radical excitations of a bound state might provide other generations \cite{Visnjic}. Another possibility is that the second and the third generation might be the bound states of the first generation $L_1$ with one and two Higgs fields \cite{Derman:1980rr} or other fields \cite{Fritzsch:1981} respectively, e.g. $\mu \sim [L_1h], ~\tau \sim [L_1 h h]$. 

There also could be fermion-string bound-states in which {\it chiral} spinon zero-modes are trapped in a vortex configuration of the chargon \cite{XC-15}. This may happen when the chargon condenses and vortex configurations appear, which is similar to the (gauged) axion-string models \cite{Callan-Harvey, XC-13, XC-14}: the spinon zero-mode is localized on the chargon vortex due to an exponential profile $
\mathcal{F}_L = \chi_{L} \, \exp \big[- \int_0^{\rho}  f(\rho') d\rho' \big], $
where $\rho$ is the spacetime polar coordinate and $f(\rho)$ are determined by the chargon and $\chi_L$ is some two-dimensional spinor (see Ref.  \cite{Callan-Harvey, XC-15, XC-13, XC-14} for details). 

There are other questions like what the binding forces would be for the bound states, how to de-gauge other theories in general and etc. We assume that some Yukawa type-interaction provide such binding force. New gauge interactions (like the hypercolor in Ref. \cite{Fritzsch:1981}) are possible, but they will destroy our spin-charge separation scenario as we do not want to have a de-gauging process followed by a new gauging process. Spacetime symmetries, especially supersymmetry may provide Yukawa-type fermion-boson interactions \cite{WB} to work as the binding forces.  Our de-gauging procedure can be generalized to other theories. As we demonstrated above for the Standard Model, the polar form or exponential parametrization of the Higgs (\ref{Ugauge}) can always be reached since there always exists a unitary gauge in the compact gauge group cases. One then can make field-decomposition as in (\ref{redef}) and identify the chargon and the spinons. However the de-gauging process of quark and color charges is still under investigation and will be given in a separate publication.   

\section*{Acknowledgement}

I thank Peter Minkowski, Harald Fritzsch, Kerson Huang, HweeBoon Low and Haibin Su for valuable discussions. This work is supported by the Institute of Advanced Studies, Nanyang Technological University, Singapore.

\bibliographystyle{ws-procs975x65}
\bibliography{ws-pro-sample}


\end{document}